\documentclass[aps,prl]{revtex4}
\usepackage{amsmath}
\usepackage{epsfig}
\usepackage{amssymb}
\usepackage{dcolumn}
\usepackage{graphicx}
\usepackage{dcolumn}

\usepackage{color}

\newcommand{\beq}{\begin{equation}}
\newcommand{\eeq}{\end{equation}}
\newcommand{\dpsi}{\delta \psi}

\newcommand{\diff}{\mathrm d}

\newcommand{\rv}{\mathbf{r}}
\newcommand{\vv}{\mathbf{v}}
\newcommand{\kv}{\mathbf{k}}

\newcommand{\F}{\mathbf{F}}

\newcommand{\dr}{{\rm d} \rv}
\newcommand{\dk}{{\rm d} \kv}

\begin{document}
\date{\today}

\title{Gaussian impurity moving through a Bose-Einstein superfluid}

\author{Florian Pinsker}

\affiliation{Clarendon Laboratory, University of Oxford, Parks Road, Oxford OX1 3PU, United Kingdom}
\email{florian.pinsker@gmail.com}

\begin{abstract}
In this paper a finite Gaussian impurity moving through an equilibrium Bose-Einstein condensate at $T= 0$ is studied. The problem can be described by a Gross-Pitaevskii equation, which is solved perturbatively. The analysis is done for systems of $2$ and $3$ spatial dimensions. 
The Bogoliubov equation solutions for the condensate perturbed by a finite impurity are calculated in the co-moving frame. 
From these solutions the total energy of the perturbed system is determined as a function of the width and the amplitude of the moving Gaussian impurity and its velocity. In addition we derive the drag force the finite sized impurity approximately experiences as it moves through the superfluid, which proves the existence of a superfluid phase for finite extensions of the impurities below the speed of sound. Finally we find that the force increases with velocity until an inflection point from which it decreases again.
\end{abstract}

\maketitle

\section{Introduction}

One fundamental property of Landau's theory of superfluidity is that below a certain critical velocity an impurity moves dissipationless through the superfluid, since excitations are energetically out of reach \cite{teo}.  In the early $20$th century  this idea has been found to be realised in Bose-Einstein condensates (BEC) occupying the ground state of a many-body system below a critical temperature \cite{uno, due, tre, quattro}.  Later Helium II was considered to represent such a state of dissipationless flow \cite{Lon}, while more recently superfluidity was confirmed for various weakly interacting and dilute gases of atoms or molecules at ultra-low temperatures forming a BEC in the nano Kelvin range \cite{first, second, third, evid, ketter, heat}. Only a few years ago the direct observation of superfluidity in terms of frictionless flow in $2$d Bose gases has been observed experimentally \cite{2D1}. Besides superfluidity has been observed in exotic quasi-Bose-Einstein condensates such as those of Exciton-Polaritons at temperatures close to the Kelvin regime or even at room temperature \cite{exi, Hugo}. Theoretically in the semiclassical mean-field regime the corresponding order parameter of the BEC describing all the atoms in the ground state, the so called condensate wave function,  is governed by the Gross-Pitaevskii equation (GPE) \cite{Pita1, Gross}. 

Utilising this GPE framework here we study the behaviour of the  superfluid by the well-established scenario of an obstacle in relative motion to the Bose-Einstein superfluid \cite{frisch, pom, win}. While a superfluid phase can be expected for low velocities above a critical velocity of relative motion between the obstacle and the superfluid a drag force arises due to the possibility of emission of elementary excitations that break down the superfluid phase \cite{pita}. For the scenario of a small impurity it was shown that the emergence of the doublet is the first excitation to break down the superfluid phase at a critical velocity below the speed of sound, while even smaller obstacles do not change the critical velocity at all and so it equals the speed of sound of the Bose-Einstein condensate \cite{diss}. In general the quantitative investigation of critical velocities in BEC can be used to probe the superfluidity and beyond criticality various additional topological phase transitions of the fluid associated with certain velocities \cite{pita, reviews}. 

In $2$ spatial dimensions for example the excitation creation  and the critical velocity of the relative motion is comparable to the situation of a rotating BEC, where below a critical rotation velocity the condensate is vortex free, while above various phases from vortex generation to giant vortices have been reported \cite{F,F1,F2,F3, F15,F16,F17}. Thus it is a matter of motion of the condensate and the form of the perturbing impurity, at which velocities of the impurity excitations are generated and what kind of excitations \cite{Pinsker, Pavloff}. Furthermore the physical dimension of the Bose-Einstein condensate sets the stage for the possible excitations a superfluid can carry, from dark solitons in $1$d to vortices in $2$d to vortex rings in $3$d \cite{Pinsker}.

The mathematical description of an impurity in a quantum gas is non-trivial and various approaches to this topic exist.  E.g. in \cite{1} the effect of a pointwise obstacle in a Fermi gas has been studied in detail using various techniques in several interaction regimes, while the works of \cite{2} focuses on the weakly-interacting and \cite{3} on the strongly-interacting regime. Furthermore we note that in particular the drag force depending on the nature of the obstacle and superfluidity has been studied in \cite{3D1} numerically for $1$D and in \cite{3D2} experimentally. In \cite{diss} an asymptotic analysis has determined the criticality of the superfluid phase transitions due to a small impurity and in \cite{pita} a perturbative approach involving Bogoliubov equations has been studied. 

Now here we follow the approach of \cite{pita} and analyse the system for a finite weakly interacting Gaussian impurity with a certain amplitude and width.  This Gaussian perturbation is moving at a constant velocity through the BEC, while thereby perturbing the condensate. We will investigate the wave generated by the perturbation and the onset and magnitude of the drag force acting upon the impurity as it reaches a critical velocity. Finally we note that our theoretical treatment includes the results presented in \cite{pita} as a limiting case, when the Gaussian becomes the Dirac delta function, i.e. a point like impurity without extension.

This paper is structured as follows. First I will determine the linear waves generated by the perturbation moving through the BEC as function of the impurity parameters for general dimensions. Secondly the energy of the perturbed system will be specified in $2$d and $3$d. Then explicit formulas for the drag force depending on dimensionality will be obtained. Finally I will discuss the results.

\section{Linear waves and energy of a small moving Gaussian impurity}
\subsection{Gross-Pitaevskii equation}

The energy of a Bose-Einstein condensate with self-interaction strength $g$, which depends on dimensionality \cite{Inge}, and which experiences the external potential $V$ is given by \cite{Inge, reviews}
\begin{equation}\label{min}
\mathcal E [ \phi ] =  \int \bigg(\frac{\hbar^2}{2m}\big| \nabla \phi \big|^2 + \frac{g}{2} |\phi|^4 +  V |\phi|^2 \bigg) \dr.
\end{equation}
While naturally GP theory is considered for  $3$d, reduction to lower dimensions occurs e.g. for strong confinement along a spatial dimension and we refer to the rigorous GP treatment of this topic in \cite{Inge}. In $3$d $g = 4 \pi \hbar^2 a /m$ and $\max_\rv V (\rv)= 4 \pi \hbar^2 b /m$ are particle-particle and particle-impurity coupling constants, with $a$ and $b$ being the respective scattering lengths \cite{pita}, while for $2$ dimensions those definitions are modified, $g_{2 {\rm d}} = \sqrt{2 \pi} \hbar^2 a/ a_z$, where $a_z = \sqrt{\hbar / m \omega_z}$ is the oscillator length \cite{Inge, pita}. 
The minimiser of \eqref{min}, say $\psi (\rv, t)$, is the so called condensate wave function, i.e. the mode in which all particles of the dilute weakly interacting gas condense \cite{reviews, Inge}. The wavefunction maps from $(\rv, t) \in \mathbb R^{d+1} \to \mathbb C$. For the sake of simplicity of notation and clarity of structure we will set $\hbar =1 = 2m$ in what follows. Now by performing a variation of the energy $\mathcal E [ \psi ]$ with respect to $\psi^*$ under the norm preserving constraint $\| \psi \|^2_2 = 1$, corresponding to the conservation of the particle number, we get the Euler-Lagrange equation for the minimiser, the Gross-Pitaevskii equation, which is given by
\begin{equation}\label{model}
 i \frac{\partial \psi}{\partial t} = - \Delta \psi + V \psi + g |\psi|^2 \psi - \mu \psi.
\end{equation} 
Here the chemical potential, i.e. the energy needed to add another particle, is $\partial \mathcal  E [ \psi ]/\partial N = \mu$ \cite{Inge}.
To model the moving Gaussian impurity we set the external potential $V=V_0 e^{- \frac{1}{2 \sigma^2} (\rv - \vv t)^2}$, which corresponds to an inserted atom or a laser beam \cite{reviews, Pavloff}. Experiments of moving obstacles in BEC have successfully shown the onset of a drag force \cite{ketter, heat}, which will be discussed theoretically in a later section.
\subsection{Linear waves}
For a single moving obstacle in any dimension the Gross-Pitaevskii energy functional is given by
\begin{equation}
\mathcal E [ \psi ] =  \int \bigg(\big| \nabla \psi \big|^2 + \frac{g}{2} |\psi|^4 + V_0 e^{- \frac{1}{2 \sigma^2} (\rv - \vv t)^2} |\psi|^2 \bigg)\dr.
\end{equation}
To derive the energy of linear waves and their form we make an ansatz of the form $\psi = \phi_0 + \delta \psi$, where $\phi_0$ represents the unperturbed part solving the GPE without potential and $\dpsi (\rv ,t )$ a small perturbation as a consequence of the Gaussian impurity. By inserting this ansatz in $\eqref{model}$ and dropping terms of order $\dpsi^2$ we get the Bogoliubov equation for the perturbation of the homogeneous system
\begin{equation}\label{lin}
 i \frac{\partial \dpsi}{\partial t} = - \Delta \dpsi + V (\phi_0 + \dpsi  )+ g \left(2 |\phi_0|^2 - \frac{\mu}{g} \right) \dpsi + g \phi_0^2 \dpsi^*.
\end{equation}
Here again we use the abbreviation for the Gaussian impurity $V=V_0 e^{- \frac{1}{2 \sigma^2} (\rv - \vv t )^2}$. Furthermore, as in \cite{pita, glad} we neglect $V \dpsi$, since we restrict our consideration to small impurities, i.e.,  $V \simeq \delta \psi$. Furthermore we will make use of the identity $\partial \dpsi(\rv - \vv t)/ \partial t = - \vv \nabla \dpsi(\rv - \vv t)$ \cite{pita, glad}. In addition we shall work in the frame moving with the impurity such that $\dpsi (\rv,t) = \dpsi (\rv - \vv t) = \dpsi (\rv')$, where we switch notations as well in the potential term. Then we drop the superscripts and consider the Fourier transform of \eqref{lin}. We use the convention $\mathcal F (f) \equiv \hat{f}(\kv) =
\int_{\mathbb R^d} f(\rv)\, e^{-i \kv \cdot \rv}\, \diff \rv$
which denotes the Fourier transform of $f(\rv) = \frac{1}{(2\pi)^d}\int_{\mathbb R^d} \hat f(\kv)\,
e^{i \kv \cdot \rv}\, \diff \kv$. So the wavefunction $\dpsi_\kv =  \int e^{-i \kv \rv } \dpsi \dr$ satisfies the Fourier transformed Bogoliubov equation given by
\begin{equation}\label{lin2}
\kv \vv \dpsi_\kv = k^2 \dpsi_k + \int e^{-i \kv \rv } V \phi_0 \dr +  g \left(2 |\phi_0|^2 - \frac{\mu}{g} \right) \dpsi_\kv + g \phi_0^2 \dpsi^*_{-\kv},
\end{equation}
where $|\kv| = k$. We now set $\phi_0 = \sqrt{n}$ and recall that $\mu = g n = c^2$, where $c$ is the speed of sound. Hence,
\begin{equation}\label{lin3}
\kv \vv \dpsi_\kv = k^2 \dpsi_\kv + \mu \left( \dpsi_\kv + \dpsi^*_{-\kv} \right) + \int e^{-i \kv \rv } V \phi_0 \dr.
\end{equation}
The remaining integral in e.g. $2$d is $2 \pi \sqrt{n} \sigma^2 V_0  e^{- \sigma^2 k^2}  \equiv f_{2 {\rm d} }(k^2)$.
Thus we consider the algebraic equation
\begin{equation}\label{lin5}
 \kv  \vv \dpsi_\kv = k^2 \dpsi_k + \mu \left( \dpsi_\kv + \dpsi^*_{-\kv} \right) + 2 \pi \sqrt{n} V_0  \sigma^2 e^{- \sigma^2 k^2},
\end{equation}
which is analytically solved by
\beq\label{solution}
\dpsi_\kv = f_{2 {\rm d} }(k^2) \frac{ \kv \vv + k^2}{(\kv \vv)^2 - k^2 (k^2 + 2 \mu )}.
\vspace{1mm}
\eeq
This solution for the linear waves generalises the solutions in \cite{pita},\cite{glad}. More precisely speaking, the Gaussian impurity generalises the Dirac delta function and so includes an additional form factor $f_{2 {\rm d} }(k^2, \sigma, V_0)$. The form factor is determined by the Fourier transform of the impurity and is for a Gaussian impurity of dimension $D$ explicitly given by
\beq
f_{{\rm D}} = (2\pi)^{D/2}  \sqrt{n} \sigma^D V_0  e^{- \frac{D}{2}\sigma^2 k^2} .
\eeq
For extensions of this type of solution to non-equilibrium systems,  which include particle spin, we refer to similar results presented in \cite{mk, mk2}.


\subsection{Energy}

\subsubsection{2d Energy of perturbed condensate}

Next we turn to the explicit form of the energy of the Gaussian obstacle moving through the homogeneous condensate. We again utilise the ansatz of the form $\psi = \phi_0 + \delta \psi$, where the perturbation $\delta \psi$ of the homogeneous system $ \phi_0$ is caused by the Gaussian impurity. So the total energy of the perturbed system is formally given by
\begin{equation}
\min_\psi \mathcal E [ \psi ] \equiv E = E_0 + E_{\text{per}}.
\end{equation}
Here the energy without perturbation is $E_0 = N g n/2$. To estimate the energy of $E_{\text{per}}$ we proceed as follows. We neglect terms of the perturbation higher than quadratic order $g \dpsi^2$, where $g$ is assumed to be small as well. Thus, we write
\beq
g |\psi |^2 \simeq g (\phi_0^2 +  \phi_0 (\dpsi+ \dpsi^*))
\eeq
\beq
g  |\psi |^4 \simeq g (\phi_0^4 +  2\phi_0^3 (\dpsi+ \dpsi^*))
\eeq
\begin{widetext}
and so the energy of the perturbation (in the co-moving frame) can be approximated by
\begin{multline}\label{per}
E'_{\text{per}} \simeq \int \left(\big| \nabla \dpsi \big|^2 + g \phi_0^3 (\dpsi+ \dpsi^*) + V(\phi_0^2 + \phi_0 (\dpsi+ \dpsi^*)) -  \mu \phi_0 (\dpsi+ \dpsi^*)  \right) \dr = \\ = \int \left(\big| \nabla \dpsi \big|^2 + V(\phi_0^2 + \phi_0 (\dpsi+ \dpsi^*)) \right) \dr =  \int \left(\big| \nabla \dpsi \big|^2 + V \phi_0 (\dpsi+ \dpsi^*) \right)\dr  + n V_0 \sigma \sqrt{2} \pi^{3/2},
\end{multline}
where we have again used $\mu = n g$ and integrated the Gaussian impurity potential in the last line. We observe that as $\sigma$ or $V_0$ increases the energy of the leading order potential energy term increases directly proportional in each of the two arguments. Let us now estimate the second term of the last line in \eqref{per}. As presented in the appendix in detail we apply the approximation $\kv \vv \ll \mu$ \cite{pita} and some algebra to obtain the explicit expression for the potential energy. It is given by
\begin{equation}\label{laa}
\int  V \phi_0 (\dpsi+ \dpsi^*)   =  \pi n (  \sigma^2 V_0 v)^2 \left(  \frac{1}{4 \mu} - e^{4 \mu \sigma^2} \sigma^2 \Gamma (0, 4 \mu \sigma^2) \right).   
\end{equation}  
\end{widetext}
The integration yields a term including the Gamma function $\Gamma(a, z) = \int^\infty_z t^{a-1} e^{-t} dt$. We note that for $\sigma \to 0$ that $\sigma^2 \Gamma (0, 4 \mu \sigma^2) \to 0$ as the chemical potential is fixed.

To calculate the kinetic energy in $\eqref{per}$ we consider it in momentum space, i.e, we obtain the quadratic dispersion,
\begin{equation}\label{kin}
\int \big| \nabla \dpsi \big|^2 \dr  =  \int k^2 \dpsi_\kv^2  \frac{\dk}{(2\pi)^2}.
\end{equation}
We remark also \cite{mk,mk2} for detailed discussions on general non-quadratic kinetic energy dispersions and note that the framework discussed here is applicable to those cases as well. While we neglect all terms of order $g \dpsi^2$, we include the kinetic energy \eqref{kin} partially in our results as it is of order $\dpsi^2$. We approximate $\dpsi_k^2$ by assuming $v$ to be small and thus we consider an expansion in $v$ up to the quadratic order, i.e.
\beq\label{mamamia}
\dpsi_\kv^2 = \frac{f_{2{\rm d}}^2 \left(k^4 + 2 k^2 \kv \vv \right)}{k^4 (k^2 + 2\mu)^2} + \mathcal O (v^4).
\eeq

Inserting $\eqref{mamamia}$ in the kinetic energy term \eqref{kin} we get by using the internal symmetries of the problem and some algebra, as presented in the appendix, the kinetic energy contribution 
\begin{equation}\label{but}
 \int k^2 \frac{f_{2{\rm d}}^2 \left(k^4 + 2 k^2 \kv \vv \right)}{k^4 (k^2 + 2\mu)^2}  \frac{\dk}{(2\pi)^2} = c \left( 
   e^{4 \mu \sigma^2} (1 + 4 \mu \sigma^2) \Gamma(0, 4 \mu \sigma^2) -1 \right),
\end{equation}  
where we have introduced the abbreviation $c= \pi n(  \sigma^2 V_0)^2 $ and used the Gamma function $\Gamma(a, z) = \int^\infty_z t^{a-1} e^{-t} dt$. We note that as $\sigma \to 0$ the r.h.s. of Eq. \ref{but} tends to zero as well, which is consistent with the homogeneous solution without perturbation while for larger values of $\sigma$ the energy is monotonically increasing. Note, however, that we consider a weakly interacting perturbation by the previously applied approximations.

Now adding all contributions from above we arrive at the formula for the total energy of the perturbed Bose-Einstein condensate in the co-moving frame, i.e.
\begin{widetext}
\begin{multline}\label{leiberl}
\frac{E'(V_0, \sigma)}{N}= \frac{gn}{2} +  \frac{n V_0 \sigma \sqrt{2} \pi^{3/2} }{N}  + \frac{\pi n (  \sigma^2 V_0 v)^2}{ N} \left(  \frac{1}{4 \mu} - e^{4 \mu \sigma^2} \sigma^2 \Gamma (0, 4 \mu \sigma^2) \right)  + \\  + \frac{ \pi n(  \sigma^2 V_0)^2}{N}  \left( 
   e^{4 \mu \sigma^2} (1 + 4 \mu \sigma^2) \Gamma(0, 4 \mu \sigma^2) -1 \right).
\end{multline}
\end{widetext}
First we note that in terms of the stationary frame the energy in the moving frame is given via a Galilean transformation $E' = E - {\bf p} \cdot \vv + \frac{1}{2} v^2$ with ${\bf p} = \int n \nabla \phi$ with $\psi = \sqrt{n} e^{i \phi}$ and ${\bf p}' = {\bf p} - \vv$. $E$ and ${\bf p}$ are the energy and momentum of the superfluid in the stationary reference frame \cite{puta}.
Further we note that as $V_0 \to 0$ or $\sigma \to 0$ we obtain $E' \to E'_0$, thus proving internal consistency.  This property particularly stems from the proportionality of those quantities to the linear waves solution \eqref{solution}.
We note that each term in this series can be identified with a term in the energy series derived for a delta potential impurity in \cite{pita} and if we let $V_0 \to \infty$ while $\sigma \to 0$ within $\eqref{per}$ the limiting case of the delta distribution is obtained. The result in \cite{pita}, however,  requires a renormalisation procedure of the coulpling constant to remove the large $k$ divergency, which is naturally avoided by the presence of the finite Gaussian potential in our proof, since it converges fast enough to zero as $k \to \infty$. Note that the potential term increases quadratically with the velocity $v$ in the moving frame \eqref{leiberl}, while there the kinetic energy is only due to the wave function modification as a consequence of the presence of the Gaussian.

\subsubsection{3d Energy of perturbed condensate}

Similarly, with calculations discussed in more detail in the appendix, we obtain the $3$d energy in the co-moving frame. It is given by
\begin{widetext}
\begin{multline}
\frac{E'_{3 \rm d}(V_0, \sigma)}{N}= \frac{gn}{2} +  \frac{n V_0 \sigma  (2\pi)^{3/2}  }{N}  + \frac{ \pi n ( \sigma^3 V_0 v)^2}{18 N} \bigg( \frac{
   \sqrt{2} e^{
    6 \mu \sigma^2} \pi (1 + 12 \mu \sigma^2) {\rm erfc}(\sqrt{6} \sqrt{\mu} \sqrt{\sigma^2} )}{
   \sqrt{\mu}}  -4 \sqrt(3 \pi) \sigma \bigg) + \\  + \frac{ \pi n (\sigma^3 V_0 )^2}{3 \sigma N} \left(2 \sqrt{ 3 \pi} (1 + 6 \mu \sigma^2) - 
 9 \sqrt{2} e^{6 \mu \sigma^2} \sqrt{\mu} \pi \sigma (1 + 4 \mu \sigma^2) {\rm erfc}(\sqrt{6} \sqrt{\mu} \sigma) \right),
\end{multline}
\end{widetext}
where we have used the notation ${\rm erfc} (z) = 1 - {\rm erf}(z)$ and ${\rm erf} (z) = \frac{2}{\pi} \int^z_0 e^{-t^2} dt$. Again the limiting case for vanishing extensions of the Gaussian reproduces the homogeneous solution, while the potential energy is quadratic in the velocity $v$. To obtain the energy in the stationary frame a Galilean transformation $E'_{3 \rm d} = E_{3 \rm d} - {\bf p} \cdot \vv + \frac{1}{2} v^2$ can be applied.

\section{Drag Force}

Next we investigate which force the impurity experiences as it moves through the condensate. To do so recall our ansatz $\psi = \phi_0 + \dpsi$ and the definition of the drag force as discussed in \cite{pita},
\beq\label{defdrag}
\F = - \int |\psi|^2 \nabla V \dr.
\eeq
Note that the drag force is directed along the velocity $\vv$. Collecting above definitions and switching into momentum space we get by neglecting $\dpsi^2$ terms 
\begin{equation}\label{here2}
\F = -2 \sigma^2 \pi V_0 \int |\phi_0 + \dpsi|^2\vec \nabla   \int e^{i \kv \rv} e^{-\sigma^2 k^2}  \frac{\dk}{(2\pi)^2} \dr   = \F_0 -2 \sigma^2 \pi V_0 \int  \int \phi_0(\dpsi  + \dpsi^*) i \kv   e^{i \kv \rv - \sigma^2 k^2}  \frac{\dk}{(2\pi)^2} \dr,
\end{equation}
where we defined $\F_0$ in the last step. It vanished due to the symmetry of integration
\begin{equation}
\F_0=2 \sigma^2 \pi V_0 n \int  \frac{e^{- \frac{x^2+y^2}{2 \sigma^2}}}{\sigma^2} \left( \vec e_x x + \vec e_y y \right) {\rm d} x {\rm d}y = {\bf 0}.
\end{equation}
The remainder in $\eqref{here2}$ for $\dpsi$ can be calculated as presented in more detail in the appendix and by taking into account the c.c., i.e., $\dpsi + \dpsi^*$. The result is a general formula for the drag force acting upon the Gaussian impurity
\beq\label{huhu}
\F = - 8 \sigma^4 \pi^2 V_0^2 n   \int   \frac{i \kv \cdot k^2 e^{- \sigma^2 k^2} }{(\kv \vv)^2 - k^2 (k^2 + 2 \mu )}   \frac{\dk}{(2\pi)^2},
\eeq
which represents a natural generalisation of the formula for the Dirac delta impurity \cite{pita}.
In the following of this section we set $c_{2 {\rm d}}  = - 8 \sigma^4 \pi^2 V_0^2 n$, which indicates the assumed $2$ spatial dimensions. To calculate \eqref{huhu} we consider the projection of the drag force $\F$ on the velocity of the obstacle $\vv$, which we denote $F_v$, and so obtain
\begin{equation}\label{integral1}
\frac{v F_v}{c_{2 {\rm d}}} =    \frac{1}{2 \pi }\int^{k_{\rm max}}_0  e^{- \sigma^2 k^2} \frac{ k^2 dk}{\sqrt{ (v^2 - c^2) - k^2}}.
\end{equation}
The integral in \eqref{integral1} has a pole analogously to the case of a Delta impurity, so that
\beq
|k| \leq k_{\rm max} = \sqrt{ v^2 -c^2},
\eeq
implying the restriction on possible momenta and the emergence of the superfluid phase for $v < c$. As shown in \cite{diss} numerically and analytically, the onset of excitation generation at the speed of sound holds for small impurities, however larger obstacles reduce the critical velocity due to early onset excitation generation, i.e. solitary waves and vortex pairs depending on the size and magnitude of the impurity. We proceed by realising that \eqref{integral1} can be solved exactly. The result is
\begin{equation}
F_v =  \frac{c_{2 {\rm d}}}{8 v}  k_{\rm max}^2 e^{-\frac{k_{\rm max}^2 \sigma^2}{2}} \cdot \left( {\rm I}_0 \left(\frac{k_{\rm max}^2 \sigma^2}{2} \right) - {\rm I}_1 \left(\frac{k_{\rm max}^2 \sigma^2}{2} \right) \right),
\end{equation}
with $I_n(z)$ being the Bessel function of the first kind. For illustration of the finiteness of the obstacle we can state the analytical upper and lower bounds to the projected force, i.e.
\beq
e^{- \sigma^2 k_{\rm max}^2}(v^2 - c^2) \leq \frac{8 \pi v F_v}{c_{2 {\rm d}}} \leq   (v^2 - c^2).
\eeq
The bounds show that in comparison with a Delta impurity the drag force is increased (due to $c_{2 {\rm d}} < 0$). Furthermore when we assume $k_{\rm max}$ to be small, we can expand the Gaussian in \eqref{integral1} and so obtain
\beq
F_v \simeq \frac{c_{2 {\rm d}}}{2 \pi v}  \left( \frac{\pi }{4 }k^2_{\rm max}  - \frac{3 \sigma^2 k^4_{\rm max}}{16}+ \frac{5 \sigma^4 k^6_{\rm max} }{64} \right),
\eeq
which to the first order term resembles the Delta impurity drag force discussed in \cite{pita}.
Note that the bounds on the $2$d Gaussian drag force do not include the effects of excitation generation, such as vortex pairs, as studied in e.g. \cite{pom, diss}, which add an additional contribution to the drag force as well as to the energy. Next we turn to the explicit form of the $3$d drag force.

\vspace{30mm}
\begin{figure}[ht]
\begin{tabular}{c}
\begin{picture}(150,0)
\put(-50,-80) {\includegraphics[scale=0.7]{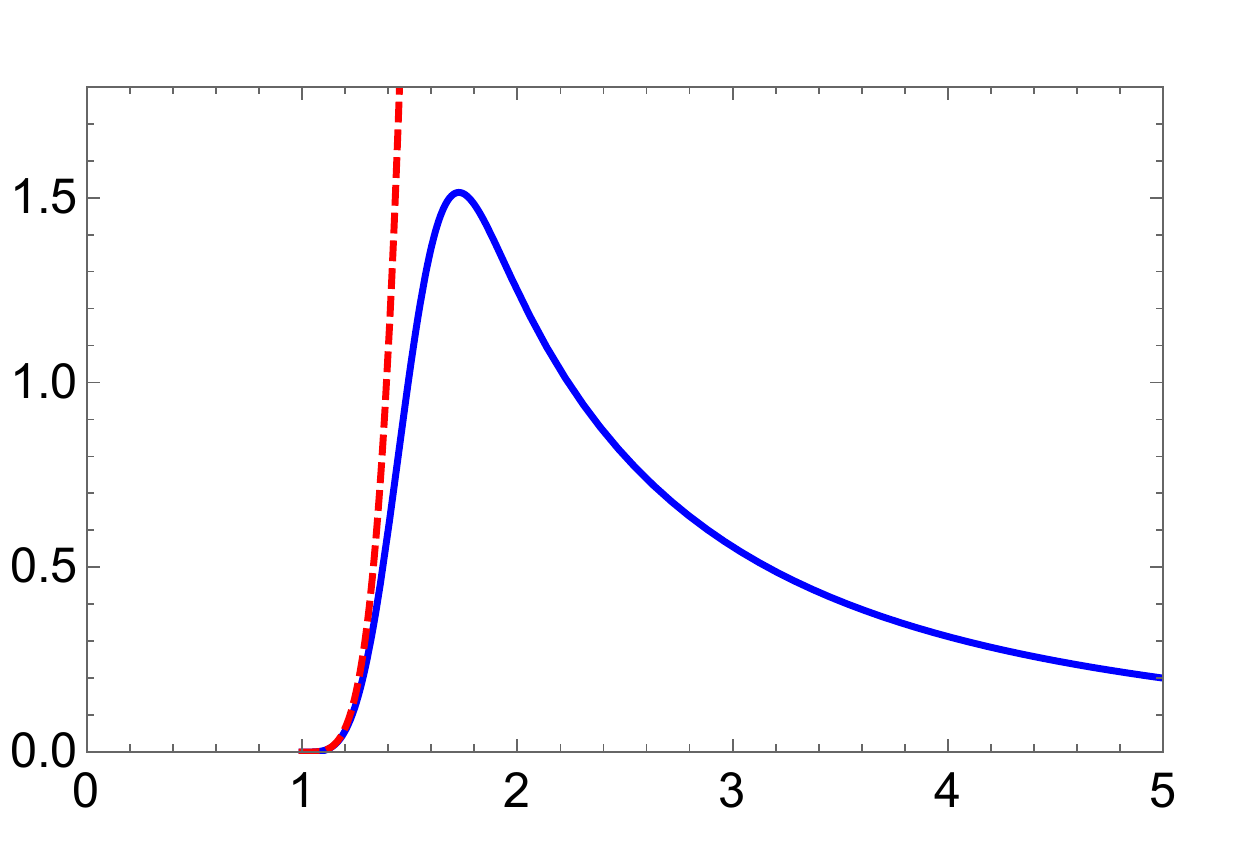} }
\put(160,50){ \textcolor{black}{}}
\put(-50,60){ \textcolor{black}{$F_v$}}
\put(70,-80){ \textcolor{black}{$v$}}
\end{picture} 
\end{tabular}

\vspace{25mm}

\caption{  Drag force projected on $\vv$ for a Gaussian (blue line) and a Dirac impurity (red line).  Numerical parameters are given in \cite{Parameters}. The intersection of the graphs with the axis at $F_v =0$ can be identified with a superfluid regime below the speed of sound.}
\label{F1}
\end{figure}
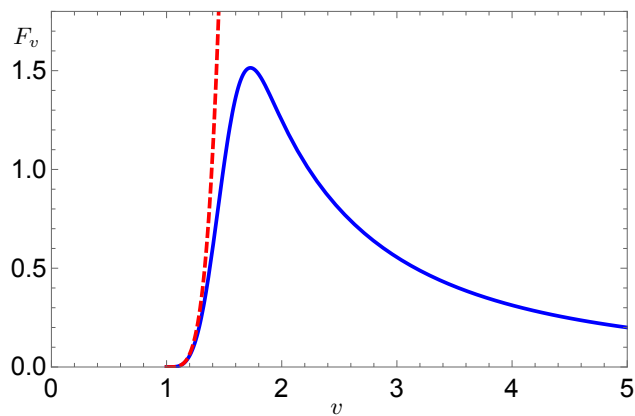

\subsection{3d drag force}

To calculate the $3$d properties of the drag force explicitly according to Landau's causality rule we may add an infinitesimal positive imaginary part to the frequency $\kv \vv$, i.e., 
\begin{equation}
\F=  c_{3 {\rm d}}   \int  e^{- \sigma^2 k^2}  \frac{i \kv \cdot k^2}{(\kv \vv + i 0)^2 - k^2 (k^2 + 2 \mu )}   \frac{\dk}{(2\pi)^3},
\end{equation}
with $c_{3 {\rm d}} = - 2((2\pi)^{3/2}  \sqrt{n} \sigma^3 V_0 )^2$. We now apply $\frac{1}{x+i 0} = P \frac{1}{x} - i \pi \delta (x)$ as in \cite{pita} where only the imaginary parts contribute due to the integration between symmetric limits and get
\begin{equation}\label{da}
\frac{\F}{c_{3 {\rm d}} }=    \int  e^{- \sigma^2 k^2}  \frac{\kv k^2 \pi}{2 \kv \vv} \bigg(\delta\left((\kv \vv ) - \sqrt{k^2 (k^2 + 2 \mu )} \right) + \delta\left((\kv \vv ) + \sqrt{k^2 (k^2 + 2 \mu )} \right) \bigg)   \frac{\dk}{(2\pi)^3}.
\end{equation}
There is no dependence on $\phi$ so the integration reduces to the following form:
\begin{equation}
\int g (\kv) \dk = \int^\infty_0 \int^\pi_0 g(k, \theta ) 2 \pi k^2 \sin \theta {\rm d} k {\rm d}\theta =  \int^\infty_0 \int^1_{-1} g(k, \cos \theta) 2 \pi k^2 {\rm d} k {\rm d}(\cos \theta).
\end{equation}
Furthermore we can utilise here the identity established in \cite{pita}, i.e.
\begin{equation}
\delta\left((\kv \vv ) \pm \sqrt{k^2 (k^2 + 2 \mu )} \right) =   \frac{1}{k v} \delta\left(\cos \theta \pm \frac{1}{k v} \sqrt{k^2 (k^2 + 2 \mu )} \right). 
\end{equation}
Here we observe that the poles in the integration over $\cos\vartheta$ appear, if the square root in the denominator is smaller than one, which again leads to the restriction on the values of momentum that contribute \cite{pita}, i.e.
\beq
|k| \leq k_{\rm max} = \sqrt{ v^2 -c^2},
\eeq
This concludes the proof of existence of a superfluid phase for a weakly interacting $3$d Gaussian obstacle.
Thus the energy dissipation takes place only if the impurity moves with a speed larger than the speed of sound \cite{pita}.
Summarising above results we obtain the projected force in direction of movement $\vv$ by multiplying it with \eqref{da} and integration, 
\begin{equation}
F_v =  c_{3 {\rm d}}  \int^{v^2 -c^2}_0 e^{- \sigma^2 k^2}  \frac{k^2 \pi}{v} \frac{1}{k v} \frac{2 \pi k^2}{(2 \pi)^3} dk   =  - 2 n (2\pi)^{3} ( \sigma V_0 )^2  \frac{1 - e^{- k_{\rm max}^2 \sigma^2} (1 + k_{\rm max}^2 \sigma^2)}{ \pi  v^2}.
\end{equation}
The drag force vanishes as $\sigma, V_0 \to 0$ showing consistency with the unperturbed system. Note that by the definition of \eqref{defdrag} we assume that $v >0$, as otherwise it is vanishing as well. Increasing the width $\sigma$ or the amplitude $V_0$ increases the effective force quadratically to the leading order. When increasing the velocity with all other parameters fixed the force monotonically increases as well to a certain value until it declines again for even larger $v$ as presented in Fig. \ref{F1}.




\section{Conclusions}

In this work first the energy of a Gaussian impurity moving through a Bose-Einstein condensate was derived utilising Bogoliubov's perturbation theory. We obtained particularly the kinetic and potential energy contribution due to the moving obstacle.  Furthermore utilising this framework it was then confirmed that for extended weakly interacting finite size impurities the motion at small velocities is still disipationless in systems of dimensionality equal or greater than $2$d up to the quadratic order. Above the speed of sound formulas for the drag force acting upon the impurity opposite the direction of motion could be specified. Physically movement with velocities larger than the speed of sound leads to a non-zero drag force due to Cherenkov radiation of phonons as previously noted for the Dirac delta impurity in \cite{pita}, which as shown here holds true for the weakly interacting finite size impurity. Furthermore the force depends on the width and amplitude of the moving obstacle in the stated analytical form and is not strictly increasing with velocity. This stands in stark contrast to the strictly increasing drag force due to infinite Dirac delta impurities reported in \cite{pita} and is a key feature of the finite obstacle. Finally we note that the analysis presented here does not include the effects of nonlinear excitations such as vortices, vortex rings, solitary waves and solitons, which add energy and cause additional drag which would add to the contributions presented here. That said, our analysis clarifies the {\it linear waves contribution} to the energy of the Bose-Einstein condensate and the drag force acting on the finite weakly interacting impurity.

\section{Acknowledgements} F.P. acknowledges financial support through his Schr\"odinger Fellowship (Austrian Science Fund (FWF): J3675) at the University of Oxford and the NQIT project (EP/M013243/1).  I would like to thank Guillaume Lang for very helpful comments.

  \newpage
  
\section{Appendix}

To calculate the kinetic energy in $\eqref{per}$ we have considered it in momentum space, i.e, 
\begin{equation}
\int \big| \nabla \dpsi \big|^2 \dr  =  \int k^2 \dpsi_\kv^2  \frac{\dk}{(2\pi)^2}.
\end{equation}
Here we used $\dpsi_\kv = \dpsi_\kv^*$ and a standard calculation in QM,
\begin{multline}
\int \left(\int e^{- i \kv' (\rv - \vv t)} \dpsi_{\kv'} \dk' \int \nabla^2 e^{i \kv (\rv - \vv t)} \dpsi_\kv  \frac{\dk}{(2\pi)^2}  \right) \dr  = \\= \int \int \int \left( e^{- i \kv' (\rv - \vv t)} \dpsi_{\kv'}   k^2 e^{i \kv (\rv - \vv t)} \dpsi_\kv   \right) \dr  \dk'  \frac{\dk}{(2\pi)^4}= \\ = \int \int \left(  \int e^{i (\kv - \kv' )\rv} \dr \cdot  \dpsi_{\kv'}   k^2  \dpsi_\kv   \right)  \dk'  \frac{\dk}{(2\pi)^4} = \\= \int \int \left(  \delta ( \kv - \kv' ) \cdot  \dpsi_{\kv'}   k^2  \dpsi_\kv   \right)  \dk'  \frac{\dk}{(2\pi)^4}= \int   k^2  \dpsi_\kv^2   \frac{\dk}{(2\pi)^2}.
\end{multline}
Inserting $\eqref{mamamia}$ in the kinetic energy term (see Eq. \ref{kin} in the main text) we get 
\begin{multline}
 \int k^2 \frac{f_{2{\rm d}}^2 \left(k^4 + 2 k^2 \kv \vv \right)}{k^4 (k^2 + 2\mu)^2}  \frac{\dk}{(2\pi)^2} =   \int k^2 \frac{f_{2{\rm d}}^2}{(k^2 + 2\mu)^2}  \frac{\dk}{(2\pi)^2}= \\ =  \frac{(2 \pi \sqrt{n} \sigma^2 V_0)^2}{(2\pi)^2} \int^\infty_0 \int^{2 \pi}_0  e^{- 2 \sigma^2 k^2} k^2 \frac{k}{(k^2 + 2\mu)^2}  dk d\phi = \\ = \pi n(  \sigma^2 V_0)^2  \left( 
   e^{4 \mu \sigma^2} (1 + 4 \mu \sigma^2) \Gamma(0, 4 \mu \sigma^2) -1 \right).
\end{multline}  
On the other hand for $3$d we obtain the kinetic energy similarly, i.e.
\begin{multline}
\int k^2 \frac{f_{3{\rm d}}^2}{(k^2 + 2\mu)^2}  \frac{\dk}{(2\pi)^3} =  \frac{(2\pi)^{3}  n (\sigma^3 V_0 )^2}{(2\pi)^3 } \int^\infty_0   e^{- 3 \sigma^2 k^2}  \frac{k^4}{(k^2 + 2\mu)^2}  (4 \pi) dk  = \\ = \frac{4 \pi n (\sigma^3 V_0 )^2}{ 12 \sigma} \bigg(2 \sqrt{ 3 \pi} (1 + 6 \mu \sigma^2) - \\ -
 9 \sqrt{2} e^{6 \mu \sigma^2} \sqrt{\mu} \pi \sigma (1 + 4 \mu \sigma^2) {\rm erfc}(\sqrt{6} \sqrt{\mu} \sigma \bigg).
\end{multline}

To calculate the energy of the potential term we  proceed as follows. Thus
\begin{multline}
\int  V \phi_0 (\dpsi+ \dpsi^*)  =  2 \sqrt{n} V_0 \int   \int\left( e^{- \frac{1}{2 \sigma^2} (x^2+ y^2)} e^{i \kv \rv } d \rv  \right) (\dpsi_k  + \dpsi^*_{-k}) \frac{\dk}{(2\pi)^2}  =  \\ = 2 \sqrt{n} V_0 \int   \int\left( e^{- \frac{1}{2 \sigma^2} (x^2+ y^2)} e^{i \kv \rv } d \rv  \right) \cdot f_{2 {\rm d} }(k^2) \frac{2 k^2}{(\kv \vv)^2 - k^2 (k^2 + 2 \mu )}  \frac{\dk}{(2\pi)^2}.
\end{multline} 
 We apply the approximation of the wave function for $\kv \vv \ll \mu$ \cite{pita}, which yields
 \beq
\int \dpsi_\kv \dk \simeq  \int f_{2 {\rm d} }(k^2) \frac{(\kv \vv)^2}{k^2 (k^2 + 2 \mu)^2} \dk  \ldots 
 \eeq
 So
\begin{equation}
  \int   \int f_{2 {\rm d} }^2 (k^2) \frac{ 2}{ (k^2 + 2 \mu )^2}  \frac{\dk}{(2\pi)^2}  = 2 \pi \frac{(2 \pi \sqrt{n} \sigma^2 V_0 v)^2}{D (2\pi)^2} \left(  \frac{1}{4 \mu} - e^{4 \mu \sigma^2} \sigma^2 \Gamma (0, 4 \mu \sigma^2) \right),
\end{equation} 
where we used the symmetry relation $\int g(k) (\kv \vv)^2 d^D k = 1/D \int g(k) (k v)^2 d^D k$ \cite{pita}. Using
\beq
f_{3 {\rm d}}= (2\pi)^{3/2}  \sqrt{n} \sigma^3 V_0  e^{- \frac{3}{2}\sigma^2 k^2} 
\eeq
we obtain the corresponding expression for the potential energy in $3$d, i.e.
\begin{equation}
  \int   \int f_{3 {\rm d} }^2 (k^2) \frac{ 2}{ (k^2 + 2 \mu )^2}  \frac{\dk}{(2\pi)^3}  =  \frac{4 \pi n ( \sigma^3 V_0 v)^2}{D 24} \bigg( \frac{
   \sqrt{2} e^{
    6 \mu \sigma^2} \pi (1 + 12 \mu \sigma^2) {\rm erfc}(\sqrt{6} \sqrt{\mu} \sqrt{\sigma^2} )}{
   \sqrt{\mu}}  -4 \sqrt(3 \pi) \sigma \bigg),
\end{equation} 
where ${\rm erfc} (z) = 1 - {\rm erf}(z)$ and ${\rm erf} (z) = \frac{2}{\pi} \int^z_0 e^{-t^2} dt$.

\subsection{Drag Force}

Furthermore to calculate the $2$d drag force we have used following:

Collecting above results and switching into momentum space we get by neglecting $\dpsi^2$ terms 
\begin{multline}\label{here}
\F = -2 \sigma^2 \pi V_0 \int |\phi_0 + \dpsi|^2 \vec \nabla   \int e^{i \kv \rv} e^{-\frac{1}{2} \sigma^2 k^2}  \frac{\dk}{(2\pi)^2} \dr\\ \simeq -2 \sigma^2 \pi V_0 \int  \int (\phi_0^2 + \phi_0(\dpsi  + \dpsi^*)) i \kv   e^{i \kv \rv} e^{-\frac{1}{2} \sigma^2 k^2}   \frac{\dk}{(2\pi)^2} \dr  \\ = \F_0 -2 \sigma^2 \pi V_0 \int  \int \phi_0(\dpsi  + \dpsi^*) i \kv   e^{i \kv \rv-\frac{1}{2} \sigma^2 k^2}  \frac{\dk \dr}{(2\pi)^2},
\end{multline}
where we defined $\F_0$ in the last step. To calculate the remainder in $\eqref{here}$ for $\dpsi$, we use the following calculation:
\begin{widetext}
\begin{multline}\label{ho}
-2 \sigma^2 \pi V_0 \int  \int \phi_0\dpsi i \kv   e^{i \kv \rv} e^{- \sigma^2 k^2}  \frac{\dk}{(2\pi)^2} d \rv = -2 \sigma^2 \pi V_0 \phi_0 \int  \int \int e^{i \rv \kv'} \dpsi_{k'}   i \kv   e^{i \kv \rv} e^{- \sigma^2 k^2}  \frac{\dk}{(2\pi)^4}  \dk' \dr = \\ =-2 \sigma^2 \pi V_0 \phi_0 \int  \int \int e^{  i(\kv' + \kv) \rv} \dpsi_{k'}   i \kv    e^{- \sigma^2 k^2}  \frac{\dk}{(2\pi)^4} \dk' \dr = -2 \sigma^2 \pi V_0 \phi_0 \int   \int \delta(\kv' + \kv) \dpsi_{k'}   i \kv    e^{- \sigma^2 k^2} \frac{\dk}{(2\pi)^4} \dk'  =\\= -2 \sigma^2 \pi V_0 \phi_0   \int  \dpsi_{- k}   i \kv    e^{- \sigma^2 k^2}  \frac{\dk}{(2\pi)^2} = - 4 \sigma^4 \pi^2 V_0^2 n   \int  e^{- \sigma^2 k^2}  \frac{- \kv \vv + k^2}{(\kv \vv)^2 - k^2 (k^2 + 2 \mu )}   i \kv \frac{\dk}{(2\pi)^2}
\end{multline}
The projection of the drag force $\F$ on the velocity of the obstacle $\vv$ is obtained via
\begin{equation}\label{integral}
\frac{v F_v}{c_{2 {\rm d}}} =    \int^\infty_0 \int^1_{-1} e^{- \sigma^2 k^2} \frac{i  2 k v \cos \theta k^2}{(kv \cos \theta)^2 - k^2 (k^2 + 2 \mu)} \frac{k}{\sqrt{1 - \cos^2 \theta}} \frac{dk d \cos \theta}{(2\pi)^2} =  \frac{1}{2 \pi} \int^{k_{\rm max}}_0  e^{- \sigma^2 k^2} \frac{ k^2 dk}{\sqrt{ (v^2 - c^2) - k^2}}.
\end{equation}
\end{widetext}



\begin{thebibliography}{10}

\bibitem{teo} L. P. Pitaevskii, Journal of Low Temperature Physics, {\bf 87}, Issue 3, pp 127-135 (1992).

\bibitem{uno} P. Kapitza, Nature {\bf 141}, 74,  (1938).

\bibitem{due} J. F. Allen, A. D. Misener, Nature {\bf 141}, 75, (1938).

\bibitem{tre} F. London, Nature {\bf 141}, 643,  (1938).

\bibitem{quattro} A. Einstein, Sitzungsber. Preuss. Akad. Wiss., 3 (1925).

\bibitem{Lon} L. Tisza, Nature {\bf 141}, 913 (1938).


\bibitem{first} M. H. Anderson \emph{et al.}, Science \textbf{269}, 198 (1995).



\bibitem{second} K. B. Davis \emph{et al.}, Phys. Rev. Lett. \textbf{75}, 3969 (1995).

\bibitem{third} C. C. Bradley  \emph{et al.},  Phys. Rev. Lett. \textbf{75}, 1687 (1995).

\bibitem{evid} C. Raman \emph{et al.},  Phys. Rev. Lett. {\bf{83}} (13), pp. 2502-2505 (1999).

\bibitem{ketter} R. Onofrio \emph{et al.}, Phys. Rev. Lett. {\bf 85}, 2228-2231 (2000).


\bibitem{heat} C. Raman \emph{et al.}, Journal of Low Temperature Physics, {\bf 122}, pp. 99 (2001).

\bibitem{2D1} R. Desbuquois, L. Chomaz, T. Yefsah, J. Léonard, J. Beugnon,
C. Weitenberg, and J. Dalibard, Nat. Phys. {\bf 8}, 645 (2012).

\bibitem{exi} A. Amo, J. Lefr\'ere, S. Pigeon, C. Adrados, C. Ciuti, I. Carusotto, R. Houdr\'e, E. Giacobino \& A. Bramati, Nature Physics {\bf 5}, 805 - 810 (2009).

\bibitem{Hugo} F. Pinsker, H. Flayac, arXiv:1310.7500, Phys. Rev. Lett. {\bf 112} (14), 140405 (2014).

\bibitem{Pita1} L. P. Pitaevskii,  Zh. Eksp. Teor. Fiz. \textbf{40}, 646 (1961); Sov. Phys. JETP \textbf{13}, 451 (1961).

\bibitem{Gross} E. P. Gross, Nuovo Cimento \textbf{20} (1961) 451; J. Math. Phys. \textbf{4} 195 (1963). 

\bibitem{frisch} T. Frisch, Y. Pomeau and S. Rica, Phys. Rev. Lett. {\bf{69}}, 1644-1647 (1992).

\bibitem{pom} C. Josserand, Y. Pomeau and S. Rica, Physica D {\bf 134}, 111-125 (1999).

\bibitem{win} T. Winiecki, J. F. McCann, and C. S. Adams, Phys. Rev. Lett. {\bf 82}, 5186-5189 (1999).



\bibitem{pita} G.E. Astrakharchik, L.P. Pitaevskii, Phys. Rev. A {\bf 70}, 013608 (2004).

\bibitem{diss} F. Pinsker, N. G. Berloff
Phys. Rev. A 89 (5), 11 (2014).

\bibitem{reviews} L. Pitaevskii, S. Stringari, \emph{Bose-Einstein condensation}, Oxford University Press, Oxford (2003).





\bibitem{F} A.L. Fetter, Phy. Rev. A \textbf{64}, 063608 (2001). 

\bibitem{F1} M. Correggi  \emph{et al.}, Phys. Rev. A \textbf{84},  053614 (2011).

\bibitem{F15} M. Correggi  \emph{et al.}, The European Physical Journal Special Topics 217 (1), 183-188 (2013).

\bibitem{F16} M. Correggi  \emph{et al.}, Journal of Physics: Conference Series 414 (1), 012034 (2013).

\bibitem{F17} M. Correggi  \emph{et al.}, Journal of Mathematical Physics 53 (9), 095203 (2012).

\bibitem{F2} M. Correggi  \emph{et al.},  J. Stat. Phys. {\bf 143}, 261--305 (2011).

\bibitem{F3} A.L. Fetter, Rev. Mod. Phys. \textbf{81}, 647--691 (2009).

\bibitem{Pinsker} F. Pinsker, {\it Excitations in superfluids of atoms and polaritons}, Department of Applied Mathematics and Theoretical Physics, University of Cambridge (2014).

\bibitem{Pavloff} N. Pavloff, Phys. Rev. A {\bf 66}, 013610 (2002).


\bibitem{1} A. Cherny, J.S. Caux and J. Brand, Front. Phys., {\bf 7}, 1, pp 54-71 (2012).

\bibitem{2} A. G. Sykes, M. J. Davis, and D. C. Roberts
Phys. Rev. Lett. {\bf 103}, 085302 (2009).

\bibitem{3} G. Lang, F. Hekking, A. Minguzzi, Phys. Rev. A {\bf 91}, 063619 (2015).


\bibitem{3D1} V. P. Singh \emph{et al.}, Phys. Rev. A {\bf 93}, 023634 (2016).

\bibitem{3D2} Weimar  \emph{et al.}, Phys. Rev. Lett. {\bf 114}, 095301 (2015).


\bibitem{Inge} Lieb E.H., Seiringer R., Solovej J.P., Yngvason J., \textit{The Mathematics of the Bose Gas and its Condensation.} Oberwolfach Seminars, {\bf 34}, Birkh\"auser, Basel, $184$pp.  (2001)

\bibitem{glad} Y.G. Gladush, L.A. Smirnov and A.M. Kamchatnov, J. Phys. B: At. Mol. Opt. Phys. {\bf{41}} 165301 (2008).

\bibitem{mk} F. Pinsker,  W. Bao, Y. Zhang, H. Ohadi, A. Dreismann, and J. J. Baumberg, Phys. Rev. B {\bf 92}, 195310  (2015).

\bibitem{mk2} F. Pinsker, X. Ruan, T. Alexander
arXiv preprint arXiv:1606.02130 (2016).


 \bibitem{puta} L. Pitaevskii and S. Stringari, {\it Bose-Einstein condensation}, Clarendon Press, Oxford (2008).

 \bibitem{Parameters} Numerical integration with parameters $c_{3 \rm d} = 40 \pi$, $\mu =0.04$, $\sigma = 1$ and $c=1$.


  \end{thebibliography}
  \end{document}